\documentclass[a4paper,aps,prd,10pt,preprintnumbers,showpacs,twocolumn,superscriptaddress,nofootinbib,amsmath,amssymb]{revtex4-1}
\usepackage{graphicx}
\usepackage[utf8]{inputenc}
\usepackage[T1]{fontenc}
\usepackage{cmap}
\usepackage{soul}

\def\imo{i}

\def\K{{\cal K}}

\begin{document}

\title{Two Regimes of Asymptotic Fall-off of a Massive Scalar Field in the Schwarzschild-de Sitter Spacetime}

\author{R. A. Konoplya}\email{roman.konoplya@gmail.com}
\affiliation{Institute of Physics and Research Centre of Theoretical Physics and Astrophysics, Faculty of Philosophy and Science, Silesian University in Opava, CZ-746 01 Opava, Czech Republic}

\begin{abstract}
The decay behavior of a massless scalar field in the Schwarzschild-de Sitter spacetime is well-known to follow an exponential law at asymptotically late times $t \rightarrow \infty$. In contrast, a massive scalar field in the asymptotically flat Schwarzschild background exhibits a decay with oscillatory (sinusoidal) tails enveloped by a power law.
We demonstrate that the asymptotic decay of a massive scalar field in the Schwarzschild-de Sitter spacetime is exponential. Specifically, if $\mu M \gg 1$, where $\mu$ and $M$ represent the mass of the field and the black hole, respectively, the exponential decay is also oscillatory. Conversely, in the regime of small $\mu M$, the decay is purely exponential without oscillations. This distinction in decay regimes underscores the fact that, for asymptotically de Sitter spacetimes, a particular branch of quasinormal modes, instead of a ``tail'', governs the decay at asymptotically late times. There are two branches of quasinormal modes for the Schwarzschild-de Sitter spacetime: the modes of an asymptotically flat black hole corrected by a non-zero $\Lambda$-term, and the modes of an empty de Sitter spacetime corrected by the presence of a black hole.
We show that the latter branch is responsible for the asymptotic decay. When $\mu M$ is small, the modes of pure de Sitter spacetime are purely imaginary (non-oscillatory), while at intermediate and large $\mu M$ they have both real and imaginary parts, what produces the two pictures of the asymptotic decay. In addition, we show that the asymptotic decay of charged and higher dimensional black hole is also exponential.
\end{abstract}

\maketitle

\section{Introduction}

The evolution of perturbations around a number of asymptotically flat black hole solutions can be conditionally divided into three stages: the initial outburst, dependent upon the initial conditions for the perturbations; the damped oscillations governed by the \textit{quasinormal modes}; and, at asymptotically late times $t \rightarrow \infty$, the power-law decay. For massless scalar and gravitational fields in the Schwarzschild background, the decay law is given by \cite{Price:1971fb}
\begin{equation}\label{asymptotictail}
\Psi \propto t^{-(2 \ell +3)}, \quad t \rightarrow \infty.
\end{equation}
J. Bičák found that the asymptotic decay in the Reissner-Nordström background is also power-law \cite{Bicak}:
\begin{equation}\label{asymptotictailRN1}
\Psi \propto t^{-(2 \ell +2)}, \quad \text{if} \quad Q < M,
\end{equation}
\begin{equation}\label{asymptotictailRN2}
\Psi \propto t^{-(\ell +2)}, \quad Q = M.
\end{equation}

For a massive field in the asymptotically flat Schwarzschild (or Reissner-Nordström) black hole spacetime, the decay at asymptotically late times is qualitatively different. It is oscillatory with a power-law envelope:
\begin{equation}\label{asymptotictailmassive}
\Psi \propto t^{-5/6}\sin(\mu t), \quad t \rightarrow \infty.
\end{equation}
This kind of oscillatory tails was also observed for a scalar \cite{Koyama:2000hj,Koyama:2001qw,Koyama:2001qw,Moderski:2001tk}, Proca \cite{Konoplya:2006gq}, and Dirac fields \cite{Jing:2004zb}, and for a massive graviton in the Randall-Sundrum models \cite{Seahra:2004fg} and massive gravity \cite{Konoplya:2023fmh}, though the law is different for brane localised black holes \cite{Dubinsky:2024jqi}.
The power-law enveloping decay rate might be slightly different at intermediate times, following the ringdown phase, but before the asymptotic fall-off. Thus, for a massive scalar field, the decay law at intermediate times is \cite{Hod:1998ra}
\begin{equation}\label{intermediatetail}
\Psi \propto t^{-\ell-3/2} \sin (\mu t),
\end{equation}
where $\ell$ is the multipole number. Observational aspects for massive fields in experiments with very long waves \cite{NANOGrav:2023gor} was discussed in \cite{Konoplya:2023fmh}.

It was shown long ago that in asymptotically de Sitter spacetimes, the asymptotic decay law for massless scalar and gravitational fields is purely exponential (non-oscillatory) \cite{Brady:1996za}. Additionally, it was known that fields of various spins in empty de Sitter space are characterized by quasinormal modes which can be found exactly \cite{Lopez-Ortega:2012xvr, Lopez-Ortega:2007vlo}. However, only recently was it demonstrated that the asymptotic purely exponential fall-off of gravitational \cite{Konoplya:2022xid} and massless scalar \cite{Konoplya:2022kld} fields in Schwarzschild-de Sitter spacetime is, in fact, the branch of (non-oscillatory) quasinormal modes of pure de Sitter space, deformed by the presence of a black hole. This new branch of modes, irrespective of the question of asymptotic decay, was first found for a massless scalar field in the Schwarzschild-de Sitter background \cite{Cardoso:2017soq}. The idea that the signal can be expanded into a set of quasinormal modes in the asymptotic regime was also elucidated in some mathematical works \cite{Dyatlov:2010hq, Dyatlov:2011jd, Dyatlov:2011zz, Hintz:2016gwb,Hintz:2021flc}, at least for some fields and ranges of parameters.

Despite the enormous number of interesting publications on the evolution of perturbations and quasinormal modes of asymptotically de Sitter black holes (see, for instance, \cite{Zhidenko:2003wq,Jansen:2017oag,Gonzalez:2022upu,Aragon:2020teq} and references therein), to the best of our knowledge, the question of quantifying and interpreting the decay of a massive field at asymptotically late times $t \rightarrow \infty$ has not been fully answered so far. Here, we will try to fill this gap and show that the quasinormal spectrum of pure de Sitter space, deformed by the presence of a black hole, governs the asymptotic decay of a massive scalar field. Unlike the massless case, there are different regimes of this decay depending on the values of $\mu M$ and $\Lambda$. Additionally, we will demonstrate that, rather unexpectedly, the WKB method is quite accurate in the regime $\mu M \gg 1$ and non-zero $\Lambda$-term because the effective potential has a single maximum between the event and cosmological horizon in this case.

The paper is organized as follows. In sec. II, we briefly review the wave equation and the method used for the analysis of the late-time decay, while secs. III and IV are devoted to the analysis of asymptotic decay in $D=4$ and $D>4$ dimensional spacetimes, respectively. In the Conclusion, we summarize the obtained results.

\section{The metric and the wave equation}

\begin{figure}
\resizebox{\linewidth}{!}{\includegraphics{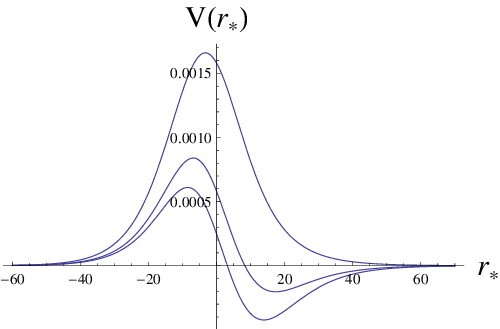}}
\caption{Effective potential for the $\ell=0$ massive scalar field in the background of the $D=4$ Schwarzschild-de Sitter black hole. From bottom to top: $\mu M = 0$ (bottom), $0.1$, and $0.2$ (top), $\Lambda M^2 = 0.1$.}\label{fig:Potential}
\end{figure}

The $D$-dimensional Schwarzschild-de Sitter black hole is described by the metric:
\begin{equation}\label{sds}
ds^2 = f(r) dt^2 - \frac{dr^2}{f(r)} - r^2 d\sigma^{D-2},
\end{equation}
where $$f(r) = 1 -\frac{2M}{r^{D-3}} - \frac{2\Lambda r^2}{(D-2)(D-1)}.$$ For the most part of this work, we will consider the $D=4$ case, where $M$ represents the black hole's mass, and $\Lambda$ is the cosmological constant.

The Klein-Gordon equation for a massive scalar field in a curved spacetime is given by:
\begin{equation}
\frac{1}{\sqrt{-g}}\partial_\mu \left(\sqrt{-g}g^{\mu \nu}\partial_\nu\Phi\right) - \mu^2 \Phi = 0,
\end{equation}
which can be reduced to the following master wave-like equation:
\begin{equation}\label{wave-equation}
\frac{\partial^2 \Psi}{\partial r_*^2} - \frac{\partial^2 \Psi}{\partial t^2} - V(r) \Psi = 0,
\end{equation}
Here, the "tortoise coordinate" $r_*$ is defined as follows:
\begin{equation}\label{tortoise}
dr_* \equiv \frac{dr}{f(r)}.
\end{equation}
The effective potential takes the form (see, for instance, \cite{Konoplya:2003ii}):
\begin{widetext}
\begin{equation}
V(r) =f(r) \left(\frac{(D-2) f'(r)}{2 r}+\frac{(D-4) (D-2)
   f(r)}{4 r^2}+\frac{\ell  (D+\ell -3)}{r^2}+\mu ^2\right).
\end{equation}
\end{widetext}
where $\ell$ is the multipole number resulting from the separation of the angular variables $\theta$ and $\phi$, and $\mu$ is the mass of the field. We will be using units $M=1$.

The effective potentials for a scalar field are shown in fig. \ref{fig:Potential}, where one can see that while for zero and small $\mu M$ the effective potential has a negative gap at a distance from the black hole, at larger $\mu M$ the gap disappears.

\section{Methods}

Quasinormal modes of an asymptotically de Sitter black holes are proper oscillation frequencies $\omega$ for which $\Psi$ satisfies the following boundary conditions:
purely ingoing wave at the event horizon and purely outgoing waves at the de Sitter horizon. Depending on the value of $\mu M$ different methods can be efficient. Here we will briefly review the three used methods: time-domain integration, WKB method and the Bernstein polynomial method.

\subsection{Time-domain integration}

Asymptotic tails could be constructed for this process via integration of the above wave-like equation in the time-domain at some fixed value of the radial coordinate. For this purpose, we use the null-cone variables $u = t - r_*$ and $v = t + r_*$ and the Gundlach-Price-Pullin discretization scheme \cite{Gundlach:1993tp}:
\begin{eqnarray}
\Psi(N) &=& \Psi(W) + \Psi(E) - \Psi(S) \nonumber \\
&& - \Delta^2V(S)\frac{\Psi(W) + \Psi(E)}{4} + \mathcal{O}(\Delta^4).\label{Discretization}
\end{eqnarray}
Here, the points are designated as follows: $N \equiv (u + \Delta, v + \Delta)$, $W \equiv (u + \Delta, v)$, $E \equiv (u, v + \Delta)$, and $S \equiv (u, v)$.

An essential aspect of computations involves the highly precise construction of the effective potential as a function of the tortoise coordinate $r_{*}$. Such precision is crucial at this stage to prevent the accumulation of numerical errors in subsequent integration steps. One approach is to integrate equation (\ref{tortoise}), derive $r^{*}$ as a function of $r$, and then use the inverse function to construct the effective potential. However, a more efficient method, which we employ here, is to treat equation (\ref{tortoise}) as a nonlinear differential equation $r'(r_{*})=f(r)$ and integrate it using the built-in function NDSolve in Mathematica, thereby obtaining the function $r(r_{*})$. This approach immediately provides us with the effective potential as a function of the tortoise coordinate.

The dominant quasinormal frequencies can be extracted from the time-domain profiles using the Prony method \cite{Konoplya:2011qq}. Although the Prony method typically cannot extract higher overtones, it accurately determines the dominant modes, except for $\ell =0$ cases where the ringing period comprises only a few oscillations. However, even in such cases, the modes governing the asymptotic decay are well determined, and it's noticeable that altering the fitting period via the Prony method to reproduce $\omega$ does not change the results for the obtained dominant frequencies. To ensure the stability of our results, we also reduce the integration step and overall precision of the procedure, verifying that neither the time-domain profile nor the extracted frequencies change significantly.

This method has been used with confirmed high accuracy in a number of works (for instance, \cite{Konoplya:2020bxa, Bolokhov:2023dxq, Abdalla:2012si, Qian:2022kaq}), so we refer the reader to these works and references therein.

\subsection{WKB method}

In the frequency domain for large and intermediate $\mu M$, we used the WKB method \cite{Schutz:1985km,Iyer:1986np,Konoplya:2003ii}, based on the expansion of the solution in the 6th order WKB series in the asymptotic regions (near the event horizon and at the de Sitter horizon) and matching them with the Taylor expansion near the peak of the effective potential.
This is possible because for $\mu M$ larger than some critical value and non-zero cosmological constant the effective potential has a single maximum and decays monotonically at the event and de Sitter horizons. For small and zero $\mu M$ the effective potential has a negative gap and there are three turning points in this case, making the WKB method much less accurate.

The semi-analytic WKB approach was first applied by Will and Schutz \cite{Schutz:1985km} to determine quasinormal frequencies. Later, the Will-Schutz formula was extended to higher orders \cite{Iyer:1986np, Konoplya:2003ii, Matyjasek:2017psv} and considerably improved by implementing the Padé approximants \cite{Matyjasek:2017psv, Hatsuda:2019eoj}.

The general form of the WKB formula has the form (see, for instance  \cite{Konoplya:2019hlu}):
\begin{eqnarray}
\omega^2&=&V_0+A_2(\K^2)+A_4(\K^2)+A_6(\K^2)+\ldots \\\nonumber
&-& \imo \K\sqrt{-2V_2}\left(1+A_3(\K^2)+A_5(\K^2)+A_7(\K^2)+\ldots\right),
\end{eqnarray}
where $\K=n+1/2$ is a half-integer. The corrections $A_k(\K^2)$, of order $k$ are polynomials of $\K^2$ with rational coefficients. These corrections depend on the values of higher derivatives of the potential $V(r)$ at the maximum. We will use here a more accurate extension of the method by Matyjasek and Opala \cite{Matyjasek:2017psv} who suggested to apply Padé approximants. We employ the sixth-order WKB method with $\tilde{m}=4,5$, where $\tilde{m}$ is defined in \cite{Matyjasek:2017psv, Konoplya:2019hlu}, because this choice is the most accurate in the Schwarzschild limit.
Application of the Padé approximant to the WKB expansion \cite{Matyjasek:2017psv} diminishes the relative error usually by quite a few times.
The higher WKB method with Padé approximants have been effectively used in a great number of recent works (for example, \cite{Matyjasek:2021xfg,Konoplya:2023ahd,Malik:2023bxc}). Therefore, we do not repeat technical aspects of this method here. For more references and details, see a review in \cite{Konoplya:2019hlu}.

\subsection{Bernstein polynomial method}

In the regime of small $\mu M$, the method based on the expansion of the solution into Bernstein polynomials \cite{Fortuna:2020obg} is an effective tool. The details of this method, when applied to asymptotically de Sitter black holes, can be found in \cite{Konoplya:2022zav}.

For asymptotically de Sitter spacetimes the purely outgoing wave is required at the de Sitter horizon.
Following the procedure suggested in \cite{Fortuna:2020obg}, we introduce a new compact coordinate as follows:
\begin{equation}\label{compact}
u\equiv\frac{\frac{1}{r}-\frac{1}{R}}{\frac{1}{r_+}-\frac{1}{R}},
\end{equation}
where $R$ is de Sitter radius and $r_{+}$ is the event horizon radius.  Then we define the new function $\psi(u)$, which is regular at $0\leq u\leq 1$, provided $\omega$ is a quasinormal frequency,
\begin{equation}\label{regularized}
\Psi(u)=u^{-\imo\Omega_c(\omega)}(1-u)^{-\imo\Omega_+(\omega)}\psi(u).
\end{equation}
Here $\Omega_c(\omega)$ and $\Omega_+(\omega)$ are obtained from the characteristic equations at the singular points, $u=0$ and $u=1$, of the wavelike equation.

In order to satisfy the quasinormal mode boundary conditions, we choose the values of $\Omega_c(\omega)$ and $\Omega_+(\omega)$, in such a way that
\begin{equation}\label{BCfix}
\frac{d\Omega_c}{d\omega~}>0,\qquad\frac{d\Omega_+}{d\omega~}>0.
\end{equation}

Then, the function $\psi(u)$ can be written as the following sum
\begin{equation}\label{Bernsteinsum}
\psi(u)=\sum_{k=0}^NC_kB_k^N(u),
\end{equation}
where
$$B_k^N(u)\equiv\frac{N!}{k!(N-k)!}u^k(1-u)^{N-k}$$
are the Bernstein polynomials.

Substituting (\ref{regularized}) into the wave equation and using a Chebyshev collocation grid of $N+1$ points,
$$u_p=\frac{1-\cos \frac{p\cdot\pi}{N}}{2}=\sin^2\frac{p\cdot\pi}{2N}, \qquad p=\overline{0,N},$$
we obtain a set of linear equations with respect to $C_k$, which has nontrivial solutions if the corresponding coefficient matrix is singular. Since the elements of the coefficient matrix are polynomials (of degree 2) of $\omega$, the problem is reduced to the eigenvalue problem of a matrix pencil (of order 2) with respect to $\omega$, which is solved numerically. Then, we calculate the coefficients $C_k$ and explicitly determine the polynomial (\ref{Bernsteinsum}), which serves as an approximation for the solution to the wave equation.

For excluding of the spurious eigenvalues, which appear due to finiteness of the polynomial basis in~(\ref{Bernsteinsum}) and do not represent true quasinormal modes, we compare both the eigenfrequencies and corresponding approximating polynomials at various $N$. First, from each set of the solutions we take the eigenvalues that differ less than the required accuracy. Then, following~\cite{Konoplya:2022xid}, for each pair of the corresponding eigenfunction, $\psi^{(1)}$ and $\psi^{(2)}$, one finds
$$1-\frac{|\langle \psi^{(1)}\;|\;\psi^{(2)} \rangle|^2}{||\psi^{(1)}||^2||\psi^{(2)}||^2}=\sin^2\alpha,$$
where $\alpha$ is the angle between the vectors $\psi^{(1)}$ and $\psi^{(2)}$ in the $L^2$-space. If all values of $\alpha$ are sufficiently small, then the obtained eigenvalues $\omega$ are reliable approximation for quasinormal frequencies, and the larger $N$ is, the better approximation we have.

\begin{figure}
\resizebox{\linewidth}{!}{\includegraphics{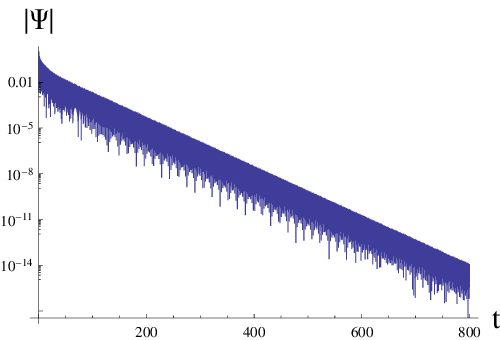}}
\caption{Time domain profile for the $\ell=0$ massive scalar field in the background of the $D=4$ Schwarzschild-de Sitter black hole: $\mu M = 10$, $\Lambda M^2 = 0.01$. The asymptotic decay is governed by the quasinormal frequency  $\omega M = 7.43036 - 0.0384134 i$. Time $t$ is measured in units of mass $M=1$.}\label{fig:timedomain1}
\end{figure}

\begin{figure}
\resizebox{\linewidth}{!}{\includegraphics{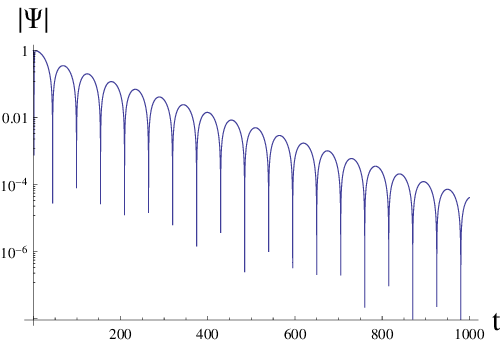}}
\caption{Time domain profile for the $\ell=0$ massive scalar field in the background of the $D=4$ Schwarzschild-de Sitter black hole: $\mu M = 1$, $\Lambda M^2 = 0.11$. The asymptotic decay is governed by the quasinormal frequency $\omega M = 0.057035 - 0.009589 i$. Time $t$ is measured in units of mass $M=1$.}\label{fig:timedomain2}
\end{figure}

\begin{figure}
\resizebox{\linewidth}{!}{\includegraphics{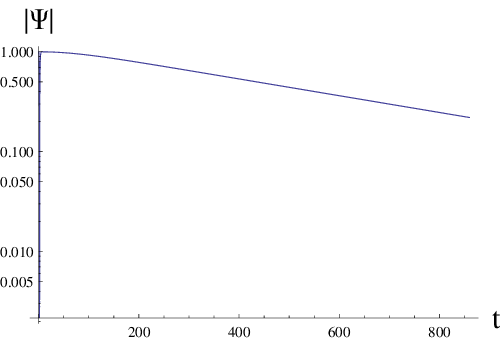}}
\caption{Time domain profile for the $\ell=0$ massive scalar field in the background of the $D=4$ Schwarzschild-de Sitter black hole: $\mu M = 0.1$, $\Lambda M^2 = 0.11$. The asymptotic decay is governed by the quasinormal frequency $\omega M = - 0.001933 i$. Time $t$ is measured in units of mass $M=1$.}\label{fig:timedomain3}
\end{figure}

\begin{figure*}
\resizebox{\linewidth}{!}{\includegraphics{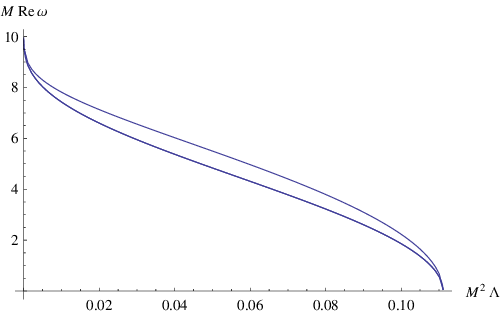}\includegraphics{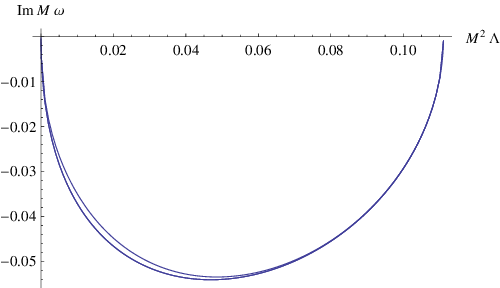}}
\caption{Quasinormal frequencies governing the asymptotic decay at $\ell=0, 1, 20$ from bottom to top for the $D=4$ Schwarzschild-de Sitter black hole, $\mu M = 10$.}\label{fig:D4}
\end{figure*}

\begin{figure*}
\resizebox{\linewidth}{!}{\includegraphics{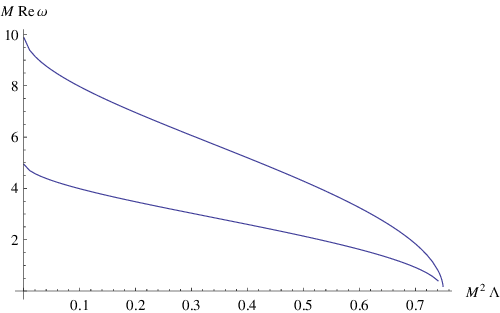}\includegraphics{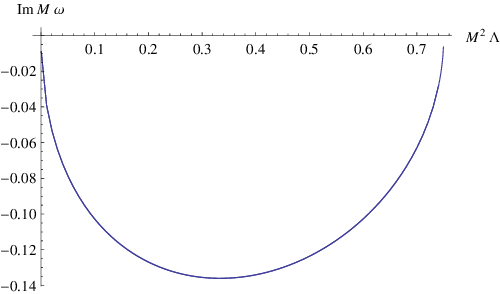}}
\caption{Real (left panel) and imaginary (right panel) parts of quasinormal frequencies governing the asymptotic decay at $\ell=0$ for $D=5$ dimensional Schwarzschild-de Sitter black hole: $\mu M = 5$ and $\mu M =10$, from bottom to top. The damping rate for different $\mu M$ is almost indistinguishable.}\label{fig:D5}
\end{figure*}

\begin{figure*}
\resizebox{\linewidth}{!}{\includegraphics{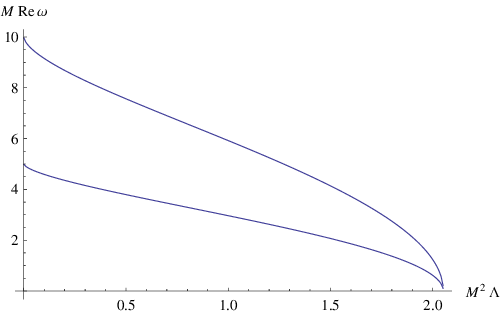}\includegraphics{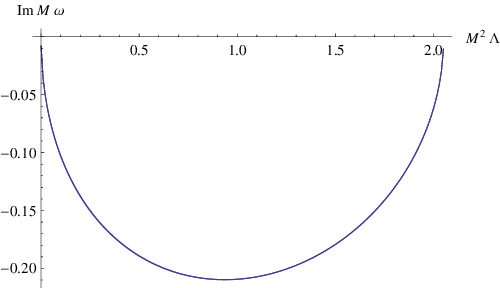}}
\caption{Real (left panel) and imaginary (right panel) parts of quasinormal frequencies governing the asymptotic decay at $\ell=0$ for $D=6$ dimensional Schwarzschild-de Sitter black hole: $\mu M = 5$ and $\mu M =10$, from bottom to top. The damping rate for different $\mu M$ is almost indistinguishable.}\label{fig:D6}
\end{figure*}

\begin{figure}
\resizebox{\linewidth}{!}{\includegraphics{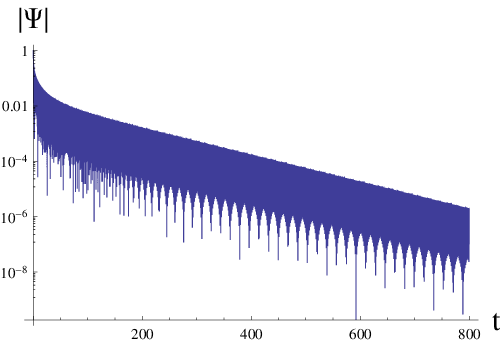}}
\caption{Time-domain profile for a $D=6$ case: $\ell=0$, $\mu M = 10$, $M =1$, $\lambda M^2 = 0.001$. At asymptotic time the frequency $\omega M = 9.95182 - 0.011187 i$ is dominant. Time $t$ is measured in units of mass $M=1$.}\label{fig:D6TD}
\end{figure}

\begin{figure}
\resizebox{\linewidth}{!}{\includegraphics{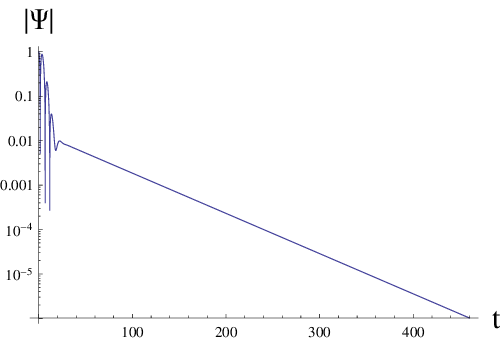}}
\caption{Time-domain profile for a $D=6$ case: $\ell=0$, $\mu M = 0.1$, $M =1$, $\Lambda M^2 = 0.1$. At asymptotic time the frequency $\omega M =   -0.0208512 i$ is dominant. Time $t$ is measured in units of mass $M=1$.}\label{fig:D6TDmu01}
\end{figure}

\begin{figure}
\resizebox{\linewidth}{!}{\includegraphics{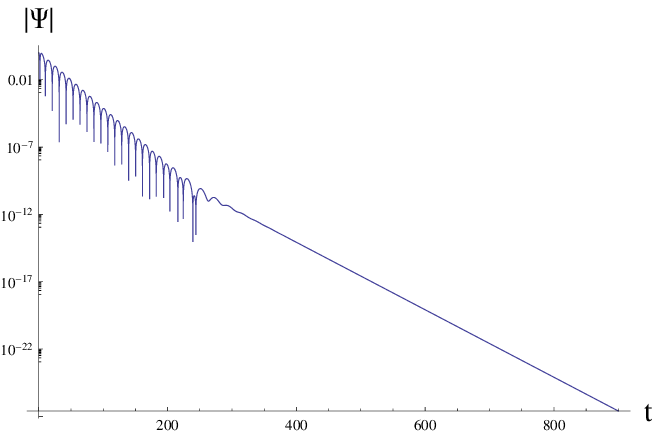}}
\caption{Time-domain profile for the $D=4$  Reissner-Nordstrom-de Sitter spacetime: $\ell=1$, $\mu M = 0$, $M = 2 Q =1$, $\Lambda M^2 = 0.001$. At asymptotic time the frequency $\omega M =   - 0.0182567  i$ is dominant. Time $t$ is measured in units of mass $M=1$.}\label{fig:RNdSTD001}
\end{figure}

\section{Decay at asymptotically lates}

\begin{table*}
\resizebox{\linewidth}{!}{
  \begin{tabular}{|c|c|c|c|c|c|c|}
     \hline
      \hline
     $\Lambda M^2$ & $\mu M =10$ (TD) & $\mu M =10$ (WKB) & $\mu M =20$ (TD) & $\mu M =20$ (WKB) & $\mu M =30$ (TD) & $\mu M =30$ (WKB)\\
    \hline
     0.01 & 7.430359 - 0.038413 i & 7.428721-0.037140 i & 14.862790 - 0.037103 i & 14.857448 - 0.037143 i & 22.304228 - 0.037053 i & 22.286173-0.037143 i\\
     0.02 & 6.598719 - 0.046644 i & 6.598255-0.046652 i & 13.200346 - 0.046623 i & 13.196609-0.046656 i &   19.807592 - 0.046561 i & 19.794941-0.046657 i\\
     0.04 & 5.372423 - 0.053709 i & 5.372165-0.053718 i & 10.746621 - 0.053688 i & 10.744606-0.053722 i &   16.123812 - 0.053660 i & 16.116986-0.053723 i\\
     0.06 & 4.308859 - 0.052767 i & 4.308733-0.052770 i & 8.6189063 - 0.052758 i & 8.617865-0.052773 i &    12.930437 - 0.052722 i & 12.926908-0.052774 i\\
     0.08 & 3.220306 - 0.045542 i & 3.220251-0.045543 i & 6.4413764 - 0.045540 i & 6.440942-0.045545 i &    9.6630035 - 0.045527 i & 9.661535-0.045545 i\\
     0.09 & 2.604175 - 0.039064 i & 2.604148-0.039065 i & 5.2089387 - 0.039062 i & 5.208710-0.039066 i &    7.8139577 - 0.039054 i & 7.813180-0.039066 i\\
     0.10 & 1.857488 - 0.029372 i & 1.857478-0.029372 i & 3.7153769 - 0.029370 i & 3.715294-0.029373 i &    5.5733168 - 0.029368 i & 5.573035-0.029373 i\\
     0.11 & 0.578221 - 0.009591 i & 0.578238-0.009590 i & 1.1565626 - 0.009591 i & 1.156595-0.009590 i &    1.7348824 - 0.009592 i & 1.734925-0.009590 i\\
     \hline
      \hline
   \end{tabular}}
 \caption{Quasinormal modes $\omega M$ governing the asymptotic decay in the regime $\mu M \gg 1$ found by the fitting of the time-domain integration at about $t/M \sim 800 $ and via the 6th order WKB method with the Padé approximants. }\label{1}
\end{table*}


\begin{table*}
\begin{tabular}{|c|c|c|c|c|}
  \hline
  \hline
  $\mu M$ & $\Lambda M^2 =0.11$ (TD)  & $\Lambda M^2 =0.11$ (Bernstein)  &  $\Lambda M^2 =0.005$ (TD) & $\Lambda M^2 =0.005$ (Bernstein) \\
 \hline
  0.02 & - 0.000070 i &-0.0000699 i &  -0.003196 i &-0.00319645 i \\
  0.05 & - 0.000445 i &-0.0004455 i &  -0.025295 i & -0.0252999 i\\
  0.07 & - 0.000894 i &-0.0008944 i & 0.037010 - 0.057920 i & 0.0373763-0.0584804 i \\
  0.1 &  - 0.001933 i &-0.0019346 i & 0.081531 - 0.052189 i &0.0818266 - 0.051095 i \\
  0.2 & 0.0064552 - 0.009680 i &0.0064598 - 0.0096870 i & 0.170307 - 0.031797 i &0.170298 - 0.0318068 i \\
  0.3 & 0.0144845 - 0.009616 i &0.0144956 - 0.0096183 i & 0.246358 - 0.026401 i &0.246341 - 0.026411 i \\
  0.5 & 0.0272925 - 0.009590 i &0.027313 - 0.00959693 i & 0.402049 - 0.0266894 i &0.40202 - 0.0266974 i \\
  \hline
  \hline
\end{tabular}
 \caption{Quasinormal modes  $\omega M$ governing the asymptotic decay for small and intermediate $\mu M$ found by the fitting of the time-domain integration at about $t/M \sim 800 $ and via the Bernstein polynomial method. }\label{2}
\end{table*}

First of all, we will consider the limit $\mu M \gg 1$, where the effective potential has a single maximum, and the WKB method can be applied for checking the time-domain integration. In a semi-logarithmic plot in Fig. \ref{fig:timedomain1}, it can be seen that at asymptotic times, the decay is oscillatory with an exponential envelope, suggesting the interpretation that the signal is represented in terms of quasinormal modes at asymptotic times. However, we can observe that a particular mode dominates in the asymptotic decay. The real oscillation frequency, given by $Re(\omega)$, is roughly proportional to $\mu$, while the damping rate almost does not change when $\mu$ is increased, what can be seen on figs. \ref{fig:D4}-\ref{fig:D6}. We also notice that once the cosmological constant goes to zero:
\begin{equation}\label{limit1}
\omega_{n=0} \rightarrow \mu, \quad \Lambda \rightarrow 0.
\end{equation}

In the limit of the extreme black hole, when the event horizon radius approaches the cosmological horizon, quasinormal frequencies tend to zero:
\begin{equation}\label{limit2}
\omega_{n=0} \rightarrow 0, \quad \Lambda \rightarrow \Lambda_{\text{ext}},
\end{equation}
which is in agreement with \cite{Cardoso:2003sw}.

To understand to which of the two branches of the Schwarzschild-de Sitter quasinormal spectrum the above modes belong, we need to remember that in the limit $\Lambda \rightarrow 0$, quasinormal modes of a massive field exhibit peculiar behavior: the damping rate of each mode diminishes when $\mu M$ is increased, such that every mode, for some particular threshold values of $\mu M$, has a vanishing damping rate. This phenomenon was called \textit{quasiresonances} \cite{Ohashi:2004wr}, and in \cite{Konoplya:2004wg}, it was shown that no such effect is possible for asymptotically de Sitter black holes. Moreover, in \cite{Zhidenko:2006rs}, it was shown that in the regime of large $\mu M$, the fundamental frequencies go to $\omega \rightarrow \mu$ only in $D>5$, while in $D=4$ and $5$, this is not the case. Notice that the quasi-resonances may not exist even in the background of  asymptotically flat black holes \cite{Zinhailo:2024jzt}.

In the limit $\mu M \ll 1$, the modes become purely imaginary (non-oscillatory), as can be seen in Table II and Fig. \ref{fig:timedomain3}, while the Schwarzschild's modes in this regime are oscillatory. Notice that similar purely exponential non-oscillatory tails are observed for lower-dimensional asymptotically de Sitter spacetime \cite{Skvortsova:2023zmj, Skvortsova:2023zca}.

In the limit $\mu M \neq 0$ and $\Lambda \rightarrow 0$, the modes do not go over into the Schwarzschild ones for a massive field, neither for large nor for small values of $\mu M$.

\begin{table*}
\begin{tabular}{|c|c|c|c|}
  \hline
  \hline
  $\mu M$ & Schwarzschild branch ($n_{S}=0$) & de Sitter branch ($n_{dS}=0$) & de Sitter branch ($n_{dS}=1$) \\
  \hline
  $0.06$ & $0.108272-0.101755 i$ & $-0.0510347 i$ & $-0.0714397 i$\\
  $0.0602$ & $0.108279-0.101735 i$ & $-0.0525371 i$ & $-0.0697937 i$ \\
  $0.0604$ & $0.108286-0.101714 i$ & $-0.0543986 i$ & $-0.0677913 i$ \\
  $0.0606$ & $0.108293-0.101693 i$ & $-0.0571186 i$ & $-0.0649330 i$ \\
  $0.0607$ & $0.108296-0.101683 i$ & $-0.0600614 i$ & $-0.0619237 i$ \\
  $0.060702$ & $0.108297-0.101682 i$ & $-0.0602298 i$ & $-0.0617540 i$ \\
  $0.060704$ & $ 0.108297-0.101682 i$ & $ -0.0604561 i$ & $-0.0615263 i$ \\
\hline
  $0.060706$ & $0.108297-0.101682 i$ & \multicolumn{2}{c|} {$ 0.0001294-0.0609905 i$} \\
  $0.06071$ & $0.108297-0.101682 i$ & \multicolumn{2}{c|} {$0.0007790-0.0609891 i$} \\
  $0.06072$ & $0.108297-0.101681 i$ & \multicolumn{2}{c|} {$0.0014411-0.0609858 i$} \\
  $0.0608$ & $0.108300-0.101672 i$ & \multicolumn{2}{c|} {$0.0037793-0.0609579 i$}  \\
  $0.0609$ & $0.108304-0.101662 i$ & \multicolumn{2}{c|} {$0.0053544-0.0609252 i$} \\
  $0.061$ & $0.108307-0.101651 i$ & \multicolumn{2}{c|} {$0.0066191-0.0608911 i$}  \\
  $0.062$ & $0.108345-0.101544 i$ & \multicolumn{2}{c|} {$0.0138321-0.0605723 i$}  \\
  $0.064$ & $0.108425-0.101325 i$ & \multicolumn{2}{c|} {$0.0220608-0.0599923 i$}  \\
  $0.068$ & $0.108613-0.100858 i$ & \multicolumn{2}{c|} {$0.0329903-0.0589618 i$}  \\
  $0.07$ & $0.108726-0.100610 i$ & \multicolumn{2}{c|} {$0.0373763-0.0584804 i$}  \\
  \hline
  \hline
\end{tabular}
 \caption{Quasinormal modes $\omega M$ found by the Bernstein polynomial method for $\Lambda M^2 =0.005$, $\ell=0$. Left column is the fundamental mode of  the Schwarzschild branch (very close to the mode of massless field in the Schwarzschild limit $\omega = 0.110455 - 0.104896 i$), while the right columns belong to the de Sitter branch, governing the asymptotic decay. Special attention is paid to the transition to oscillatory regime which occurs at $\mu M \approx 0.0607$. Here $n_{S}$ and $n_{dS}$ are numbers of overtones in the Schwarzschild and de Sitter branches respectively. At the transition two purely imaginary modes merges to form a pair of complex conjugate modes with opposite signs of $Re (\omega)$. }\label{2}
\end{table*}

When $\Lambda \neq 0$ and $\mu M \rightarrow 0$ from Table II, we can see that the frequencies governing the asymptotic decay do not transition to the quasinormal modes of the massless scalar field in the Schwarzschild-de Sitter background.

On the other hand, in the limit of a vanishing black hole mass $M \rightarrow 0$, the quasinormal modes of the empty de Sitter spacetime can be found exactly in the following form \cite{Lopez-Ortega:2012xvr, Lopez-Ortega:2007vlo}:
\begin{equation}\label{exact1}
i \omega_{n} R = \ell + 2n + \frac{D-1}{2} \pm \sqrt{\frac{(D-1)^2}{4} - \mu^2 R^2},
\end{equation}
for
\begin{equation}
\frac{(D-1)^2}{4} > \mu^2 R^2,
\end{equation}
and as follows:
\begin{equation}\label{exact2}
i \omega_{n} R = \ell + 2n + \frac{D-1}{2} \pm i \sqrt{\mu^2 R^2-\frac{(D-1)^2}{4}},
\end{equation}
for
\begin{equation}
\frac{(D-1)^2}{4} < \mu^2 R^2.
\end{equation}
Here, $R=\sqrt{3/\Lambda}$ is the de Sitter radius.

One can notice that in the case of small $\mu M$, the quasinormal modes of the pure de Sitter space are purely imaginary, which is in concordance with the limit $\mu \rightarrow 0$ for which non-oscillatory exponential modes dominate in the asymptotic decay \cite{Konoplya:2022xid}.

We can observe that for the fundamental quasinormal mode of the empty de Sitter spacetime with $\mu = 0.05$ and $\Lambda M^2 = 0.005$, $\omega_{n=0} =  - 0.0258819 i$, while for $\mu M = 0.05$ of Schwarzschild-de Sitter spacetime, according to Table II, we have $\omega_{n=0} = - 0.025295 i$. This signifies that for small $\mu M$, the quasinormal modes of black holes tend to the modes of pure de Sitter space.

To determine which branch of modes, either de Sitter or Schwarzschild, governs the asymptotic decay at intermediate $\mu M$, we calculate the dominant quasinormal modes of both branches for fixed $\Lambda M^2$ and various $\mu M$, from $\mu M=0.06$ to $\mu M=0.07$. This way we can see in detail the transition to an oscillatory regime. The results are summarized in Table III. It is evident from the table that the fundamental mode of the Schwarzschild branch remains almost unchanged, whereas the de Sitter mode transitions gradually from a purely non-oscillatory state (at zero mass) to an oscillatory regime. During this transition two purely imaginary modes merge to a pair of complex conjugate modes with opposite signs of $Re (\omega)$. The same takes place for a modes of empty de Sitter spacetime, as can be seen from  eq. \ref{exact1}. Notice also that the transition to the oscillatory regime occurs in pure de Sitter space at $\mu \approx 0.061237$ which is very close to the point of transition in Schwrazschild-de Sitter  $\mu \approx 0.0607$ ($M=1$).

Overall, we conclude that quasinormal frequencies dominating in the asymptotic decay of a massive scalar field in the $D=4$ Schwarzschild-de Sitter background are non-perturbative in $\Lambda$. In other words, the branch of modes representing the empty de Sitter spacetime but deformed by the introduction of the black hole dominates at asymptotic times.

In the regime of large $\mu M$, the quasinormal modes that dominate during the asymptotic decay can be found analytically by using the higher-order WKB expansion and expansion in terms of the inverse field's mass $1/\mu$. This approach is similar in spirit with the $1/\ell$ expansion in \cite{Konoplya:2023moy}. Using the designation of \cite{Konoplya:2023moy}:
\begin{equation}
K = n + \frac{1}{2},
\end{equation}
and introducing, for convenience and compactness of the resultant analytic expressions, a new quantity $\sigma$:
\begin{equation}
\sigma = (9 M^2 \Lambda)^{1/6},
\end{equation}
we have the position of the maximum of the effective potential $r_{0}$ in the following form:
\begin{equation}
r_{0} = \frac{3 M}{\sigma ^2}+\frac{\left(\sigma ^2-1\right)\left(\ell ^2+\ell +\sigma ^2\right)}{3 M \mu^2}+O\left(\frac{1}{\mu}\right)^4.
\end{equation}

Then, we use the above expansion in terms of $1/\mu$ when applying the 6th order WKB formula \cite{Schutz:1985km, Iyer:1986np, Konoplya:2003ii}
\begin{equation}\label{eq:11}
   \frac{i (\omega^2 - V_{0})}{\sqrt{-2 V_{0}^{\prime \prime}}}
   - \sum_{i=2}^{i=6} \lambda_{i} =n+\frac{1}{2}.
\end{equation}
Here $\lambda_{i}$ is  the $i$-th order WKB correction term, $n=0, 1, 2, \cdots $ is the overtone number,  $V_{0}$ and $V_{0}^{\prime \prime}$ is the value of the effective potential in its maximum. The WKB corrections $\lambda_{i}$ depend on derivatives of orders up to $2 i$ of the effective potential in its peak and explicit form of the corrections can be found in  \cite{Schutz:1985km, Iyer:1986np, Konoplya:2003ii}.
This gives us the analytic approximate value for the quasinormal modes
\begin{widetext}
$$\omega_{n} = \mu  \sqrt{1-\sigma ^2}-\frac{i K \sigma ^3 \sqrt{1-\sigma
   ^2}}{3 M}- \frac{\sigma ^4 \sqrt{1-\sigma ^2} \left(-72
   \ell ^2-72 \ell +12 K^2 \left(\sigma ^2-1\right)+29
   \sigma ^2-11\right)}{1296 M^2 \mu }+$$
\begin{equation}
\frac{i K \sigma ^5
   \left(1-\sigma ^2\right)^{3/2} \left(864 \ell ^2+864 \ell
   +76 K^2 \left(\sigma ^2-1\right)+865 \sigma
   ^2+167\right)}{46656 M^3 \mu
   ^2}+O\left(\frac{1}{\mu }\right)^3
   \end{equation}
\end{widetext}
The above analytic expression is in very good agreement with the data obtained in Table I. For example, $\mu M = 20$, $\Lambda M^2 = 0.08$, gives
$\omega_{n=0} M = 6.44094 - 0.0455447 i$, while the time-domain integration result is $\omega_{n=0} M = 6.4413764 - 0.045540 i$, which means that the relative error is much less than one percent.

\vspace{4mm}
\section{Higher dimensional black holes}

While four-dimensional black holes are our primary focus, the study of perturbations in higher-dimensional black holes has its own significance.
First and foremost, the higher-dimensional (Tangherlini) generalization of the Schwarzschild solution to the vacuum Einstein equations describes a black hole within the framework of extra-dimensional gravity, provided that the size of the extra dimensions is much smaller than the radius of the event horizon \cite{Arkani-Hamed:1998jmv,Randall:1999ee,Kanti:2004nr}. Within this context, numerous papers have delved into the exploration of perturbations and quasinormal modes of such black holes (see, for instance \cite{Kanti:2005xa,Abdalla:2006qj} and references therein).

Simultaneously, gauge/gravity duality \cite{Maldacena:1997re} implies that the quasinormal modes of $D+1$-dimensional black holes may be interpreted as the poles of the retarded Green function in $D$-dimensional field theory at finite temperature, which corresponds to the Hawking temperature of the horizon \cite{Horowitz:1999jd,Kovtun:2004de,Son:2007vk}. Although this application primarily concerns asymptotically AdS black holes, exploration of the correspondence between asymptotically de Sitter spacetimes and Conformal Field Theory has also been undertaken \cite{Strominger:2001pn}. Thus, understanding the quasinormal modes of higher-dimensional black holes has multiple implications, spanning from higher-dimensional gravity to gauge/gravity duality.

The Green functions in even and odd $D$ for asymptotically flat black holes are qualitatively different; so is the asymptotic decay of fields because the power-law tails are contributions from the branch cut integral in the complex frequency plane. The asymptotic decay of massive fields in Schwarzschild-de Sitter spacetime, at least in $D=4$ as we have seen in the previous section, is governed by quasinormal modes instead of power-law tails. The natural question then is what happens for higher dimensions, especially taking into account that the limit $\omega \rightarrow \mu, \quad \Lambda \rightarrow 0, \quad \mu M \gg1$ coincides for $D>5$ with the quasinormal modes of the Schwarzschild black hole \cite{Zhidenko:2006rs}.

From Figs. \ref{fig:D5} and \ref{fig:D6}, we see that the same limits given by eqs. \ref{limit1} and \ref{limit2} take place in higher dimensions. However, for $D>5$, the limit $\Lambda \rightarrow 0$ coincides with the fundamental mode of the asymptotically flat Schwarzschild black hole \cite{Zhidenko:2006rs}. The typical time-domain profiles for $D=6$ are shown in Fig. \ref{fig:D6TD} and \ref{fig:D6TDmu01}. There one can see that the two regimes of asymptotic decay (oscillatory or purely exponential) takes place, depending on the value of $\mu M$ in accordance with the general formula for quasinormal modes of empty de Sitter spacetime given by eqs. \ref{exact1} and \ref{exact2}.

Here we observe that the quasinormal modes of $D=5$-dimensional asymptotically de Sitter black holes are perturbative in $\mu$ and non-perturbative in $\Lambda, $ which is similar to the $D=4$ case. On the contrary, for $D \geq 6$, the situation is more intriguing because at least the fundamental mode might be perturbative in $\Lambda$. Indeed, when $\Lambda \neq 0,$ but $\Lambda \rightarrow 0$ and $\mu M \gg 1$, the limits, given by Eqs. \ref{limit1} and \ref{limit2}, again take place:
\begin{equation}
\omega_{n=0} \rightarrow \mu, \quad \mu M \gg 1, \Lambda \neq 0,
\end{equation}
which is the same limit for $\Lambda = 0$.

The analytical formula for quasinormal modes in the regime of large $\mu M$ can be derived in the same way as in previous section for higher $D$, though this time it is more convenient to choose, for $D=5$,
\begin{equation}
\sigma = (9 \Lambda M^{7})^{\frac{1}{4}}.
\end{equation}
Then, we obtain the position of the potential's peak at

\begin{widetext}
\begin{equation}
r_{0}=-\frac{3 \left(\sqrt{2} M^2\right)}{\sigma }+\frac{\left(9
   M^3-2 \sigma ^2\right) \left(M^3 \left(4 \ell ^2+8 \ell
   +3\right)+2 \sigma ^2\right)}{48 \sqrt{2} M^5 \mu ^2
   \sigma }+O\left(\frac{1}{\mu }\right)^3,
\end{equation}
and the analytic expression for the quasinormal modes:
\begin{equation}
\omega_{n} =\frac{1}{3} \mu  \sqrt{9-\frac{2 \sigma ^2}{M^3}}-\frac{i
   \sqrt{2} K \sigma ^2 \sqrt{\left(9 M^3-2 \sigma
   ^2\right)^3}}{243 M^8-54 M^5 \sigma ^2}+\frac{\sigma ^2
   \sqrt{9-\frac{2 \sigma ^2}{M^3}} \left(3 M^3 (\ell
   +1)^2-\sigma ^2\right)}{324 M^7 \mu
   }+O\left(\frac{1}{\mu }\right)^2
\end{equation}

For $D=6$ we choose
\begin{equation}
\sigma = (30^ 4 \Lambda M^{4})^{\frac{1}{5}}.
\end{equation}
Then, the position of the maximum of the effective potential is
\begin{equation}
r_{0} = \frac{30 M}{\sigma }-\frac{\left(5400 M^2-\sigma ^3\right)
   \left(1350 M^2 \left(\ell ^2+3 \ell +2\right)+\sigma
   ^3\right)}{121500 \mu ^2 \left(M^3 \sigma
   ^2\right)}+O\left(\frac{1}{\mu }\right)^3.
\end{equation}
The quasinormal modes are
$$ \omega_{n} =\frac{1}{30} \mu  \sqrt{900-\frac{\sigma ^3}{6 M^2}}-\frac{i
   K \sigma ^{5/2} \sqrt{\left(5400 M^2-\sigma
   ^3\right)^3}}{54000 \sqrt{3} M^3 \left(5400 M^2-\sigma
   ^3\right)}+$$
\begin{equation}
\frac{\sigma ^2 \sqrt{900-\frac{\sigma ^3}{6
   M^2}} \left(1350 M^2 \left(36 \ell ^2+108 \ell -12
   K^2+85\right)+\left(3 K^2-28\right) \sigma
   ^3\right)}{2624400000 M^4 \mu }+O\left(\frac{1}{\mu
   }\right)^2.
\end{equation}

\end{widetext}
Analytic approximations for $\omega_{n}$ can be obtained for higher $D$ in a similar fashion. However, the higher $D$, the more cumbersome is the resultant analytical formula. From the above formulas we can see that again $\omega \rightarrow \mu$, when $\Lambda \rightarrow 0$ ($\sigma \rightarrow 0$).

\section{Remarks on two recent works}

During the preparation of this paper for publication, another work \cite{Correa:2024xki} appeared on arXiv, which also addresses the asymptotic decay of a massive scalar field in an asymptotically de Sitter background. However, it is noteworthy that the work in question does not establish any connections with the quasinormal modes of empty de Sitter spacetime, which constitutes the main focus of our findings presented here.

After the initial version of this paper appeared on arXiv, I became aware of the work by R. Fontana et al. \cite{Fontana:2020syy}. In contrast to \cite{Correa:2024xki}, Fontana et al. associated quasinormal modes of asymptotically de Sitter black holes with the modes of empty de Sitter spacetime. However, in their significant work, quasinormal modes were not linked to the asymptotic decay, and furthermore, while their time-domain integration shown in fig. 6 of \cite{Fontana:2020syy} suggests that at late times the decay is exponential, self-contradictory, in fig. 7 of \cite{Fontana:2020syy} the late time tails are power-law ones even for a massless scalar field.
We believe that such a power-law tail as in fig. 7 of \cite{Fontana:2020syy} is due to chosen very small value of $\Lambda M^2 \sim 10^{-6}$, so that integration with high precision is necessary to distinguish tiny $\Lambda$ from asymptotically flat spacetime.
On the contrary to \cite{Fontana:2020syy} we always find exponential tails (either oscillatory or not) at asymptotic times, which are consistent with a general observation made in \cite{Brady:1996za,Konoplya:2022xid, Konoplya:2022kld}. For this purpose we added here the time-domain profile (a semi-logarithmic plot fig. \ref{fig:RNdSTD001}) for a charged black hole with the same charge and mass as in \cite{Fontana:2020syy}, but for a much larger value of $\Lambda M^2 = 10^{-3}$. There we can see, that the asymptotic decay is clearly exponential.

\section{Conclusions}

While perturbations and quasinormal modes of Schwarzschild-de Sitter spacetime have been extensively studied over the past two decades, the question of the asymptotic ($t \rightarrow \infty$) decay of massive fields in this simple background has remained unclear. The asymptotic decay exhibits qualitative differences depending on whether the spacetime is asymptotically flat or de Sitter, and whether the field under consideration is massive or massless.

In this work, we demonstrate that the asymptotic decay of the massive scalar field in $D=4$ and $D=5$ Schwarzschild-de Sitter spacetimes is governed by a particular branch of quasinormal modes: the modes of empty de Sitter spacetime deformed by the introduction of a black hole. This finding explains the peculiar behavior of the asymptotic decay, which is oscillatory at large $\mu M$ and purely exponential at small $\mu M$. Further studies in the case of $D \geq 6$ would be desirable. While the overall picture of asymptotic decay in $D \neq 6$ is similar, in the regime $\Lambda \rightarrow 0$ and large $\mu M$, the asymptotically flat limit is reproduced for the fundamental mode. This limit coincides with the limit of pure de Sitter modes when $\Lambda$ is small but non-zero.

In a similar vein, our work could be extended to other asymptotically de Sitter black holes in various alternative theories of gravity. In the regime of large $\mu M$, we anticipate that the WKB method will yield sufficiently accurate values of quasinormal modes governing the asymptotic decay.

This extension may hold intriguing applications, particularly considering that the dimensionless product of masses, $\mu M$, is equivalent to $\mu M/m_{P}^2$, where $m_{P}$ denotes the Planck mass. For an electron, for instance, this implies that $\mu M \sim 10$ for black holes of mass $M \sim 10^{15} kg$, representing the regime of large $\mu M$ corresponding to massive Standard Model particles and black holes spanning from primordial to large astrophysical scales.

\acknowledgments
The author would like to thank useful discussions with A. Zhidenko and S. Dyatlov. I would like to thank R. Fontana for letting me know about his work \cite{Fontana:2020syy}.


\begin{thebibliography}{80}


\bibitem{Price:1971fb}
R.~H.~Price,
Phys. Rev. D \textbf{5}, 2419-2438 (1972)


\bibitem{Bicak}
Bičák, J. "Gravitational collapse with charge and small asymmetries I. Scalar perturbations."
Gen Relat Gravit 3, 331–349 (1972).

\bibitem{Koyama:2000hj}
H.~Koyama and A.~Tomimatsu,
Phys. Rev. D \textbf{63}, 064032 (2001)
doi:10.1103/PhysRevD.63.064032
[arXiv:gr-qc/0012022 [gr-qc]];
\bibitem{Koyama:2001ee}
Phys. Rev. D \textbf{64}, 044014 (2001)
doi:10.1103/PhysRevD.64.044014
[arXiv:gr-qc/0103086 [gr-qc]];
\bibitem{Koyama:2001qw}
Phys. Rev. D \textbf{65}, 084031 (2002)
doi:10.1103/PhysRevD.65.084031
[arXiv:gr-qc/0112075 [gr-qc]].

\bibitem{Moderski:2001tk}
R.~Moderski and M.~Rogatko,
Phys. Rev. D \textbf{64}, 044024 (2001)
doi:10.1103/PhysRevD.64.044024
[arXiv:gr-qc/0105056 [gr-qc]].


\bibitem{Konoplya:2006gq}
R.~A.~Konoplya, A.~Zhidenko and C.~Molina,
Phys. Rev. D \textbf{75}, 084004 (2007)
doi:10.1103/PhysRevD.75.084004
[arXiv:gr-qc/0602047 [gr-qc]].

\bibitem{Jing:2004zb}
J.~Jing,
Phys. Rev. D \textbf{72}, 027501 (2005)
doi:10.1103/PhysRevD.72.027501
[arXiv:gr-qc/0408090 [gr-qc]].

\bibitem{Seahra:2004fg}
S.~S.~Seahra, C.~Clarkson and R.~Maartens,
Phys. Rev. Lett. \textbf{94}, 121302 (2005)
doi:10.1103/PhysRevLett.94.121302
[arXiv:gr-qc/0408032 [gr-qc]].

\bibitem{Konoplya:2023fmh}
R.~A.~Konoplya and A.~Zhidenko,
[arXiv:2307.01110 [gr-qc]].


\bibitem{Dubinsky:2024jqi}
A.~Dubinsky,
[arXiv:2403.01883 [gr-qc]].

\bibitem{Hod:1998ra}
S.~Hod and T.~Piran,
Phys. Rev. D \textbf{58}, 044018 (1998)
doi:10.1103/PhysRevD.58.044018
[arXiv:gr-qc/9801059 [gr-qc]].


\bibitem{NANOGrav:2023gor}
G.~Agazie \textit{et al.} [NANOGrav],
Astrophys. J. Lett. \textbf{951}, no.1, L8 (2023)
doi:10.3847/2041-8213/acdac6
[arXiv:2306.16213 [astro-ph.HE]].






\bibitem{Brady:1996za}
P.~R.~Brady, C.~M.~Chambers, W.~Krivan and P.~Laguna,
Phys. Rev. D \textbf{55}, 7538-7545 (1997)
doi:10.1103/PhysRevD.55.7538
[arXiv:gr-qc/9611056 [gr-qc]].


\bibitem{Lopez-Ortega:2012xvr}
A.~Lopez-Ortega,
Gen. Rel. Grav. \textbf{44}, 2387-2400 (2012)
doi:10.1007/s10714-012-1398-4
[arXiv:1207.6791 [gr-qc]].

\bibitem{Lopez-Ortega:2007vlo}
A.~Lopez-Ortega,
Gen. Rel. Grav. \textbf{39}, 1011-1029 (2007)
doi:10.1007/s10714-007-0435-1
[arXiv:0704.2468 [gr-qc]].

\bibitem{Konoplya:2022xid}
R.~A.~Konoplya and A.~Zhidenko,
Phys. Rev. D \textbf{106}, no.12, 124004 (2022)
doi:10.1103/PhysRevD.106.124004
[arXiv:2209.12058 [gr-qc]].

\bibitem{Konoplya:2022kld}
R.~A.~Konoplya and A.~Zhidenko,
JCAP \textbf{11}, 028 (2022)
doi:10.1088/1475-7516/2022/11/028
[arXiv:2210.04314 [gr-qc]].

\bibitem{Cardoso:2017soq}
V.~Cardoso, J.~L.~Costa, K.~Destounis, P.~Hintz and A.~Jansen,
Phys. Rev. Lett. \textbf{120}, no.3, 031103 (2018)
doi:10.1103/PhysRevLett.120.031103
[arXiv:1711.10502 [gr-qc]].



\bibitem{Dyatlov:2010hq}
S.~Dyatlov,
Commun. Math. Phys. \textbf{306}, 119-163 (2011)
doi:10.1007/s00220-011-1286-x
[arXiv:1003.6128 [math.AP]].

\bibitem{Dyatlov:2011jd}
S.~Dyatlov,
Annales Henri Poincare \textbf{13}, 1101-1166 (2012)
doi:10.1007/s00023-012-0159-y
[arXiv:1101.1260 [math.AP]].


\bibitem{Dyatlov:2011zz}
S.~Dyatlov,
Math. Res. Lett. \textbf{18}, 1023-1035 (2011)
doi:10.4310/MRL.2011.v18.n5.a19
[arXiv:1010.5201 [math.AP]].

\bibitem{Hintz:2016gwb}
P.~Hintz and A.~Vasy,
doi:10.4310/acta.2018.v220.n1.a1
[arXiv:1606.04014 [math.DG]].

\bibitem{Hintz:2021flc}
P.~Hintz and Y.~Xie,
Phys. Rev. D \textbf{104}, no.6, 064037 (2021)
doi:10.1103/PhysRevD.104.064037
[arXiv:2104.11810 [gr-qc]].



\bibitem{Zhidenko:2003wq}
A.~Zhidenko,
Class. Quant. Grav. \textbf{21}, 273-280 (2004)
doi:10.1088/0264-9381/21/1/019
[arXiv:gr-qc/0307012 [gr-qc]].

\bibitem{Jansen:2017oag}
A.~Jansen,
Eur. Phys. J. Plus \textbf{132}, no.12, 546 (2017)
doi:10.1140/epjp/i2017-11825-9
[arXiv:1709.09178 [gr-qc]].

\bibitem{Gonzalez:2022upu}
P.~A.~Gonz\'alez, E.~Papantonopoulos, J.~Saavedra and Y.~V\'asquez,
JHEP \textbf{06}, 150 (2022)
doi:10.1007/JHEP06(2022)150
[arXiv:2204.01570 [gr-qc]].

\bibitem{Aragon:2020teq}
A.~Arag\'on, R.~B\'ecar, P.~A.~Gonz\'alez and Y.~V\'asquez,
Phys. Rev. D \textbf{103}, no.6, 064006 (2021)
doi:10.1103/PhysRevD.103.064006
[arXiv:2009.09436 [gr-qc]].



\bibitem{Gundlach:1993tp}
C.~Gundlach, R.~H.~Price and J.~Pullin,
Phys. Rev. D \textbf{49}, 883-889 (1994)
doi:10.1103/PhysRevD.49.883
[arXiv:gr-qc/9307009 [gr-qc]].

\bibitem{Konoplya:2011qq}
R.~A.~Konoplya and A.~Zhidenko,
Rev. Mod. Phys. \textbf{83}, 793-836 (2011)
doi:10.1103/RevModPhys.83.793
[arXiv:1102.4014 [gr-qc]].

\bibitem{Konoplya:2020bxa}
R.~A.~Konoplya and A.~F.~Zinhailo,
Eur. Phys. J. C \textbf{80}, no.11, 1049 (2020)
doi:10.1140/epjc/s10052-020-08639-8
[arXiv:2003.01188 [gr-qc]].

\bibitem{Bolokhov:2023dxq}
S.~V.~Bolokhov,
[arXiv:2310.12326 [gr-qc]].


\bibitem{Abdalla:2012si}
E.~Abdalla, O.~P.~F.~Piedra, F.~S.~Nu\~nez and J.~de Oliveira,
Phys. Rev. D \textbf{88}, no.6, 064035 (2013)
doi:10.1103/PhysRevD.88.064035
[arXiv:1211.3390 [gr-qc]].

\bibitem{Qian:2022kaq}
W.~L.~Qian, K.~Lin, C.~Y.~Shao, B.~Wang and R.~H.~Yue,
Eur. Phys. J. C \textbf{82}, no.10, 931 (2022)
doi:10.1140/epjc/s10052-022-10910-z
[arXiv:2203.04477 [gr-qc]].


\bibitem{Schutz:1985km}
B.~F.~Schutz and C.~M.~Will,
Astrophys. J. Lett. \textbf{291}, L33-L36 (1985)

\bibitem{Iyer:1986np}
S.~Iyer and C.~M.~Will,
Phys. Rev. D \textbf{35}, 3621 (1987)


\bibitem{Konoplya:2003ii}
R.~A.~Konoplya,
Phys. Rev. D \textbf{68}, 024018 (2003)
doi:10.1103/PhysRevD.68.024018
[arXiv:gr-qc/0303052 [gr-qc]].

\bibitem{Matyjasek:2017psv}
J.~Matyjasek and M.~Opala,
Phys. Rev. D \textbf{96}, no.2, 024011 (2017)
doi:10.1103/PhysRevD.96.024011
[arXiv:1704.00361 [gr-qc]].

\bibitem{Hatsuda:2019eoj}
Y.~Hatsuda,
Phys. Rev. D \textbf{101}, no.2, 024008 (2020)
doi:10.1103/PhysRevD.101.024008
[arXiv:1906.07232 [gr-qc]].

\bibitem{Konoplya:2023ahd}
R.~A.~Konoplya, D.~Ovchinnikov and B.~Ahmedov,
Phys. Rev. D \textbf{108}, no.10, 104054 (2023)
doi:10.1103/PhysRevD.108.104054
[arXiv:2307.10801 [gr-qc]].

\bibitem{Matyjasek:2021xfg}
J.~Matyjasek,
Phys. Rev. D \textbf{104}, no.8, 084066 (2021)
doi:10.1103/PhysRevD.104.084066
[arXiv:2107.04815 [gr-qc]].


\bibitem{Malik:2023bxc}
Z.~Malik,
[arXiv:2308.10412 [gr-qc]].



\bibitem{Konoplya:2019hlu}
R.~A.~Konoplya, A.~Zhidenko and A.~F.~Zinhailo,
Class. Quant. Grav. \textbf{36}, 155002 (2019)
doi:10.1088/1361-6382/ab2e25
[arXiv:1904.10333 [gr-qc]].

\bibitem{Fortuna:2020obg}
S.~Fortuna and I.~Vega,
Eur. Phys. J. C \textbf{83}, no.12, 1170 (2023)
doi:10.1140/epjc/s10052-023-12350-9
[arXiv:2003.06232 [gr-qc]].

\bibitem{Konoplya:2022zav}
R.~A.~Konoplya and A.~Zhidenko,
Phys. Rev. D \textbf{107}, no.4, 044009 (2023)
doi:10.1103/PhysRevD.107.044009
[arXiv:2211.02997 [gr-qc]].


\bibitem{Cardoso:2003sw}
V.~Cardoso and J.~P.~S.~Lemos,
Phys. Rev. D \textbf{67}, 084020 (2003)
doi:10.1103/PhysRevD.67.084020
[arXiv:gr-qc/0301078 [gr-qc]].

\bibitem{Ohashi:2004wr}
A.~Ohashi and M.~a.~Sakagami,
Class. Quant. Grav. \textbf{21}, 3973-3984 (2004)
doi:10.1088/0264-9381/21/16/010
[arXiv:gr-qc/0407009 [gr-qc]].





\bibitem{Konoplya:2004wg}
R.~A.~Konoplya and A.~V.~Zhidenko,
Phys. Lett. B \textbf{609}, 377-384 (2005)
doi:10.1016/j.physletb.2005.01.078
[arXiv:gr-qc/0411059 [gr-qc]].

\bibitem{Zhidenko:2006rs}
A.~Zhidenko,
Phys. Rev. D \textbf{74}, 064017 (2006)
doi:10.1103/PhysRevD.74.064017
[arXiv:gr-qc/0607133 [gr-qc]].


\bibitem{Zinhailo:2024jzt}
A.~F.~Zinhailo,
[arXiv:2403.06867 [gr-qc]].



\bibitem{Skvortsova:2023zmj}
M.~Skvortsova,
[arXiv:2311.11650 [gr-qc]].

\bibitem{Skvortsova:2023zca}
M.~Skvortsova,
[arXiv:2311.02729 [gr-qc]].



\bibitem{Konoplya:2023moy}
R.~A.~Konoplya and A.~Zhidenko,
Class. Quant. Grav. \textbf{40}, no.24, 245005 (2023)
doi:10.1088/1361-6382/ad0a52
[arXiv:2309.02560 [gr-qc]].



\bibitem{Arkani-Hamed:1998jmv}
N.~Arkani-Hamed, S.~Dimopoulos and G.~R.~Dvali,
Phys. Lett. B \textbf{429}, 263-272 (1998)
doi:10.1016/S0370-2693(98)00466-3
[arXiv:hep-ph/9803315 [hep-ph]].

\bibitem{Randall:1999ee}
L.~Randall and R.~Sundrum,
Phys. Rev. Lett. \textbf{83}, 3370-3373 (1999)
doi:10.1103/PhysRevLett.83.3370
[arXiv:hep-ph/9905221 [hep-ph]].

\bibitem{Kanti:2004nr}
P.~Kanti,
Int. J. Mod. Phys. A \textbf{19}, 4899-4951 (2004)
doi:10.1142/S0217751X04018324
[arXiv:hep-ph/0402168 [hep-ph]].


\bibitem{Kanti:2005xa}
P.~Kanti and R.~A.~Konoplya,
Phys. Rev. D \textbf{73}, 044002 (2006)
doi:10.1103/PhysRevD.73.044002
[arXiv:hep-th/0512257 [hep-th]].

\bibitem{Abdalla:2006qj}
E.~Abdalla, B.~Cuadros-Melgar, A.~B.~Pavan and C.~Molina,
Nucl. Phys. B \textbf{752}, 40-59 (2006)
doi:10.1016/j.nuclphysb.2006.06.017
[arXiv:gr-qc/0604033 [gr-qc]].

\bibitem{Maldacena:1997re}
J.~M.~Maldacena,
Adv. Theor. Math. Phys. \textbf{2}, 231-252 (1998)
doi:10.4310/ATMP.1998.v2.n2.a1
[arXiv:hep-th/9711200 [hep-th]].

\bibitem{Horowitz:1999jd}
G.~T.~Horowitz and V.~E.~Hubeny,
Phys. Rev. D \textbf{62}, 024027 (2000)
doi:10.1103/PhysRevD.62.024027
[arXiv:hep-th/9909056 [hep-th]].

\bibitem{Kovtun:2004de}
P.~Kovtun, D.~T.~Son and A.~O.~Starinets,
Phys. Rev. Lett. \textbf{94}, 111601 (2005)
doi:10.1103/PhysRevLett.94.111601
[arXiv:hep-th/0405231 [hep-th]].

\bibitem{Son:2007vk}
D.~T.~Son and A.~O.~Starinets,
Ann. Rev. Nucl. Part. Sci. \textbf{57}, 95-118 (2007)
doi:10.1146/annurev.nucl.57.090506.123120
[arXiv:0704.0240 [hep-th]].

\bibitem{Strominger:2001pn}
A.~Strominger,
JHEP \textbf{10}, 034 (2001)
doi:10.1088/1126-6708/2001/10/034
[arXiv:hep-th/0106113 [hep-th]].


\bibitem{Correa:2024xki}
M.~M.~Corr\^ea, C.~F.~B.~Macedo and J.~L.~Rosa,
[arXiv:2401.15156 [gr-qc]].

\bibitem{Fontana:2020syy}
R.~D.~B.~Fontana, P.~A.~Gonz\'alez, E.~Papantonopoulos and Y.~V\'asquez,
Phys. Rev. D \textbf{103}, no.6, 064005 (2021)
doi:10.1103/PhysRevD.103.064005
[arXiv:2011.10620 [gr-qc]].




\end{thebibliography}
\end{document}